\documentclass{article}
\usepackage{spconf,amsmath,graphicx}
\usepackage{booktabs}
\usepackage{makecell}
\usepackage{multirow}
\usepackage{easyReview}
\usepackage{balance}
\usepackage{url}
\usepackage[subtle]{savetrees} 
\usepackage{pgfplots}
\usepackage[style=ieee, doi=false, isbn=false, url=false, eprint=false, related=false, maxnames=5, minnames=1, date=year]{biblatex}
\usepackage{xurl}
\AtEveryBibitem{\clearlist{language}}
\AtEveryBibitem{\clearfield{note}}
\AtEveryBibitem{\clearfield{address}}
\AtEveryBibitem{\clearlist{address}}
\AtEveryBibitem{\clearlist{location}}
\AtEveryBibitem{\clearfield{location}}
\AtEveryBibitem{\clearfield{publisher}}
\AtEveryBibitem{\clearlist{publisher}}
\AtEveryBibitem{\clearlist{pages}}
\AtEveryBibitem{\clearfield{pages}
    \clearfield{urlyear}
    \clearfield{urlmonth}}
\title{Voice Anonymization for all -\\ Bias Evaluation of the Voice Privacy Challenge Baseline Systems}
\name{Author(s) Name(s)}
\address{Author Affiliation(s)}
\twoauthors
  {Anna Leschanowsky}
	{Fraunhofer IIS\\
	Am Wolfsmantel 33, 91058 Erlangen, Germany}
  {Ünal Ege Gaznepoglu, Nils Peters\thanks{The International Audio Laboratories Erlangen are a joint institution of the Friedrich-Alexander-Universität Erlangen-Nürnberg and Fraunhofer IIS.}}
	{International Audio Laboratories Erlangen\\
	Am Wolfsmantel 33, 91058 Erlangen, Germany}

\begin{document}
%
\maketitle
\begin{abstract}
In an age of voice-enabled technology, voice anonymization offers a solution to protect people's privacy, provided these systems work equally well across subgroups. This study investigates bias in voice anonymization systems within the context of the Voice Privacy Challenge. We curate a novel benchmark dataset to assess performance disparities among speaker subgroups based on sex and dialect. We analyze the impact of three anonymization systems and attack models on speaker subgroup bias and reveal significant performance variations. Notably, subgroup bias intensifies with advanced attacker capabilities, emphasizing the challenge of achieving equal performance across all subgroups. Our study highlights the need for inclusive benchmark datasets and comprehensive evaluation strategies that address subgroup bias in voice anonymization.
\end{abstract}
\section{Introduction}
\label{sec:intro}

With the growing popularity of voice-enabled devices, voice privacy has become a pressing issue. The Voice Privacy Challenge (VPC) has fostered the development of voice anonymization systems that conceal a speaker's identity while leaving the linguistic content and paralinguistic attributes unchanged~\cite{tomashenkoVoicePrivacy2022Challenge2022}. These systems can be assessed for privacy and utility, e.g., using automatic speaker verification (ASV) and automatic speech recognition (ASR) respectively. At the same time, bias and discrimination are well documented for ASR and ASV systems~\cite{koeneckeRacialDisparitiesAutomated2020, limaEmpiricalAnalysisBias2019, tatmanEffectsTalkerDialect2017, martinUnderstandingRacialDisparities2020, toussaintDesignGuidelinesInclusive2022,hutiriBiasAutomatedSpeaker2022a, fenu21_interspeech, HajaviFairness2023, EstevezFairness2023}. Moreover, the organizers of the VPC have acknowledged performance differences across speaker genders, highlighting the need for further investigation~\cite{tomashenkoVoicePrivacy2020Results}.

Considering prior research and VPC results~\cite{tomashenkoVoicePrivacy2020Results}, it is likely that evaluation systems for voice anonymization algorithms also exhibit bias. Yet, unbiased evaluation is crucial to ensure that voice anonymization techniques perform well for any speaker subgroup. Moreover, data-driven approaches for anonymization can introduce inherent bias. However, due to the interplay between anonymization and evaluation systems, detecting bias becomes extremely challenging. Our paper presents an in-depth study on bias in voice anonymization. Guided by established bias evaluation guidelines, we create a benchmark dataset, focusing on speaker groups identified by sex and dialect, and evaluate bias across three anonymization systems and attack models, i.e. ignorant, lazy-informed and semi-informed attacker~\cite{tomashenkoVoicePrivacy2022Challenge2022}. We demonstrate how bias evaluation can inform system development and uncover attack models. To our knowledge, this is the first analysis of this kind for voice anonymization systems. 

\section{Voice Privacy Challenge (VPC)}
\label{sec:VPC}

The VPC has provided three baseline systems, i.e. one system based purely on signal processing techniques (B2) and two systems based on x-vector modification and distinct speech synthesis components (B1.a and B1.b)~\cite{snyderXVectorsRobustDNN2018, patino21_interspeech}. Our analysis focuses on the newer baseline systems B1.b and B2 from VPC 2022. Anonymization using B1.b is performed in three steps, i.e. extraction of x-vector, pitch and bottleneck features, x-vector anonymization and speech synthesis, most of which require training data sets. Importantly, the x-vector anonymization relies on a set of candidate x-vectors drawn from a speaker pool. To select an appropriate pseudo-speaker, several design choices exist, e.g. distance metric, proximity, gender selection~\cite{srivastavaSpeakerAnonymizationRepresentation}. In this study, we focus on the 
proximity parameter's influence on bias due to its significant impact on privacy~\cite{srivastavaSpeakerAnonymizationRepresentation}, while keeping other parameters constant. We focus on two proximity criteria for baseline B1.b, i.e. farthest and nearest, where candidate x-vectors are selected based on their distance from the original x-vector. 
The evaluation of the voice anonymization systems is based on the equal error rate (EER) of the ASV functioning as the primary privacy metric and on the word error rate (WER) of the ASR serving as a utility metric. Thus, to ensure privacy preservation while allowing for downstream tasks, anonymization systems should achieve a high EER and low WER. For more information, see~\cite{tomashenkoVoicePrivacy2022Challenge2022}.

\section{Benchmarks for Bias Evaluation}
\label{sec:TIMITBenchmark}

Popular benchmark corpora like Librispeech~\cite{panayotovLibrispeechASRCorpus2015} for ASR and VoxCeleb~\cite{NagraniVoxceleb19} for ASV can lack representative linguistic diversity~\cite{rusti2023benchmark}. Ideally, ASR benchmark corpora should predict the robustness of the system w.r.t. linguistic properties and system requirements~\cite{aksenovaHowMightWe2021} but are limited due to complex linguistic properties of speech. Few datasets exist that representatively cover certain single aspects, such as regional language variation (e.g., TIMIT~\cite{garofolojohns.TIMITAcousticPhoneticContinuous1993}), sociolects (e.g., Corpus of Regional African American Language (CORAAL)~\cite{farringtonCorpusRegionalAfrican2021}), or speech impairments (e.g., Whitaker Corpus~\cite{deller1993whitaker}). Recent research has stimulated a reevaluation of current benchmark corpora and proposed guidelines to develop more inclusive and representative alternatives~\cite{aksenovaHowMightWe2021, toussaintDesignGuidelinesInclusive2022}. In particular, it was shown that benchmarks for ASV require not only inclusion at the speaker but utterance pair level~\cite{FenuExploring2020, Fenu_2020_Improving,toussaintDesignGuidelinesInclusive2022}.

\subsection{Shortcomings of Evaluation Datasets of the VPC}

The VPC relies on the LibriSpeech test-clean subset and a subset of VCTK to evaluate voice anonymization systems~\cite{tomashenkoVoicePrivacy2022Challenge2022}. LibriSpeech test-clean consists of 5.4~h of read English speech from 20 female and 20 male speakers~\cite{panayotovLibrispeechASRCorpus2015}. VCTK-test evaluation set consists of read English speech from 15 female and 15 male speakers~\cite{tomashenkoVoicePrivacy2022Challenge2022}. 

To evaluate subgroup bias, annotated datasets including labels on gender, accent or other characteristics are required. Yet,  LibriSpeech provides only sex labels, limiting subgroup bias evaluation. Similarly, the VCTK-test set covers only English accents and was therefore deemed unsuitable. 
Further, both datasets fall short of best practices for constructing inclusive benchmark datasets for ASV at utterance  level~\cite{toussaintDesignGuidelinesInclusive2022}. 

\subsection{Benchmark dataset based on TIMIT}

Given these limitations, it becomes imperative to establish a suitable benchmark dataset for evaluating bias in voice anonymization systems. Previous studies on bias evaluation for ASV have relied on datasets like VoxCeleb or FairVoice~\cite{Fenu_2020_Improving, hutiriBiasAutomatedSpeaker2022a}. However, these datasets either serve as training data for the VPC or are not publicly available. As the VPC relies on the Kaldi toolkit\footnote{https://kaldi-asr.org/}, we use the TIMIT corpus as our base dataset which represents linguistic diversity and comes with a Kaldi recipe. TIMIT encompasses 630 speakers across eight dialects of American English, however, has varying speaker counts per dialect, ranging from 11 to 79 per gender and dialect subgroup~\cite{garofolojohns.TIMITAcousticPhoneticContinuous1993}. To ensure subgroup balance and account for one missing female speaker data, we selected ten male and ten female speakers from each dialect resulting in a total of 160 speakers. Additionally, according to guidelines by~\cite{toussaintDesignGuidelinesInclusive2022} for constructing inclusive ASV benchmarks, we address the following key requirements:

\noindent
\textbf{1. Equal number of utterance pairs} 
Limited by the number of ten recordings per speaker, we randomly select five enrollment utterances per speaker and use the remaining recordings to construct trial utterance pairs. Thus, ensuring that there is no overlap between enrollment and trial utterances. Therefore, our benchmark consists of ten utterance pairs per speaker, i.e., five same and five different speaker pairs, resulting in 800 same speaker and 800 different speaker trials. 

\noindent
\textbf{2. Difficulty grading}
Comparing utterances of the same speaker is more difficult the more different they are, e.g., different recording and different background noise, while comparing utterances of different speakers is more difficult the more similar they are, e.g., same gender and accent. 
We consider the difficulty of utterance pairs by selecting the least similar ones for same-speaker utterance pairs and the most similar ones, i.e. same sex and dialect, for different-speaker utterance pairs.

\noindent
\textbf{3. Robustness}
To ensure robustness in the evaluation, \cite{toussaintDesignGuidelinesInclusive2022} recommend having at least 500 different speaker utterance pairs. However, due to the limited TIMIT dataset size, achieving this threshold while maintaining balance on the utterance level is not possible. To compensate, we compare two randomly generated dataset variations, which we made publicly available\footnote{\url{https://github.com/AnnaLesch/bias_evaluation_TIMIT_benchmark/}}.

\section{Bias Evaluation}
\label{sec:biasevaluation}

\subsection{Speaker Verification}
\label{sec:biasasv}

\begin{figure*}[ht]
    \centering   \resizebox{0.8\linewidth}{0.6\linewidth}{\input{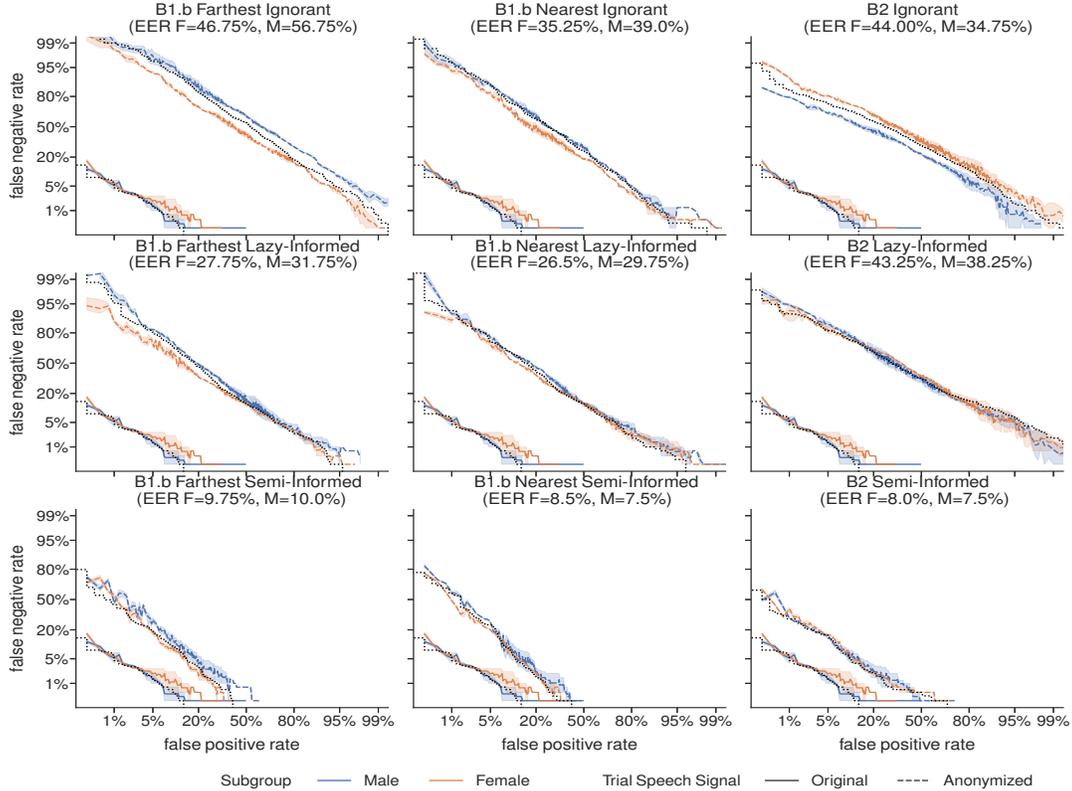}}
    \caption{DET curves disaggregated for female and male speakers across different attack models and anonymization systems together with the disaggregated results for the original ASV system without anonymization. Overall performance across subgroups is indicated by the black dotted lines. In addition, we show the respective EER values obtained for each subgroup.}
    \label{fig:subplot_gender_differences}
\end{figure*}

      

We use the python library \textit{bt4vt}\footnote{https://pypi.org/project/bt4vt/} to assess bias in speaker verification. Several types of bias can arise during model building and implementation, i.e. aggregation, learning, evaluation and deployment bias~\cite{hutiriBiasAutomatedSpeaker2022a}. While the former two focus on aspects of model building, the latter two are concerned with implementation. 
Therefore, to analyze differences between anonymization systems, we restrict evaluation to aggregation and learning bias. Further, by having constructed a suitable benchmark, we hope to mitigate evaluation bias. 

We evaluate aggregation bias by exploring the impact of sex and accent differences on anonymization system performance. To explore learning bias, we consider modelling choices, i.e., different anonymization systems and pseudo-speaker design choices. First, we plot disaggregated detection error trade-off (DET) curves to visualize the trade-off between false positive and false negative rates for speaker subgroups as they allow for deeper insights into disparities than gender-dependent EER results. In Figure~\ref{fig:subplot_gender_differences}, we present DET curves disaggregated for female and male speakers across different attack models and anonymization systems together with the results of the original ASV system without anonymization in each subplot. As we computed results for two benchmark dataset variations, we plot mean curves with 95\% confidence intervals. Thus, DET curves below the black dotted line - representing overall performance - perform better than average and curves above the dotted line perform worse than average. For ASV evaluation, we usually focus on curves above the dotted line and their corresponding subgroups, while for anonymization systems, we are interested in curves below the dotted line as they indicate a higher likelihood of recognition and thus, lower protection by anonymization. We make the following observations: In the case of an ignorant attacker, the data-driven baseline B1.b performs worse for females, while the signal processing-based baseline B2 exhibits the opposite trend. Considering that the original ASV system shows little disparities towards females, it becomes clear that the data-driven baseline reverses aggregation bias of the ASV system while the signal processing-based baseline exaggerates it. Moreover, the influence of proximity on bias is evident, with farthest proximity showing significant differences between females and males. 
Finally, bias effects seem to diminish with attacker capabilities but show robustness across benchmark dataset variations. Yet, small variations persist with the farthest proximity exhibiting the largest discrepancy. 

While attributing differences in performance to a specific module in the anonymization pipeline is challenging, differences based on proximity may be linked to an unbalanced pseudo-speaker pool. The VPC relies on the LibriTTS train-other-500 for this pool, which shows gender distribution skewing towards males both at the speaker (560 female vs. 600 male ) and utterance level (14155 female vs. 16357 male). This imbalance may result in more distant male pseudo-speakers in the farthest proximity scenario, possibly resulting in a decreased likelihood of re-identification. 

Similarly, we evaluated DET curves disaggregated for dialect regions and sex. While subgroup performance varied significantly, it was not only influenced by the anonymization system but also by the evaluated benchmark dataset variation. 


\begin{figure}[!t]
    \centering
    \resizebox{0.9\linewidth}{0.9\linewidth}{\input{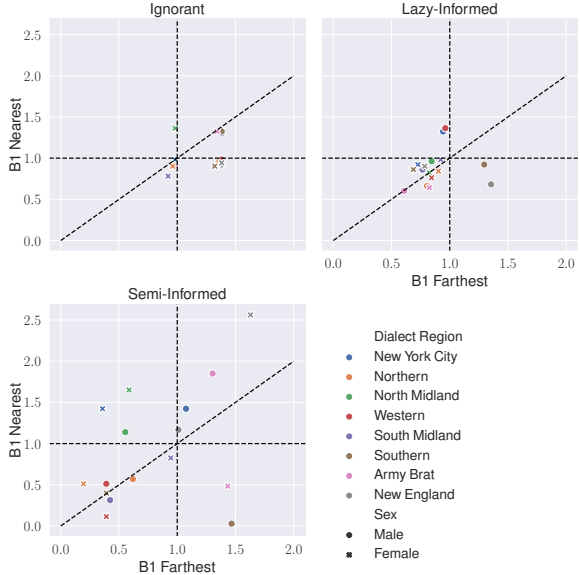}}
    \caption{Subgroup bias for the two design modifications of the baseline B1.b for the three attack models when evaluated on one of the benchmark dataset variations. On the diagonal, subgroup bias is equal for the two design choices.}
    \label{fig:farthest_nearest}
\end{figure}

For further analysis, we computed subgroup bias as defined in~\cite{hutiriBiasAutomatedSpeaker2022a}. This ratio indicates subgroup performance relative to the overall system with values greater than 1 indicating subgroup bias worse than average and values smaller than 1 suggesting performance above average. Thus, in the case of voice anonymization, subgroups with values smaller than 1 show a higher likelihood of re-identification. Importantly, we found that subgroup bias did not correlate for benchmark dataset variations with correlation values below 0.5 for anonymization systems and attack models. As previous research has highlighted the high variability in DET curves for datasets with limited utterance pairs per speaker and subgroup~\cite{toussaintDesignGuidelinesInclusive2022}, our observations are likely a result of the small benchmark dataset size. However, our results show that evaluating dataset variations can enhance evaluation reliability by preventing misinterpretations. Nevertheless, a notable trend emerged. For illustration, we show subgroup bias for both variations of baseline B1.b when evaluated on one of the benchmark datasets in Figure~\ref{fig:farthest_nearest}. It reveals increasing disparities between groups as attacker capabilities improve - an observation that is consistent across anonymization systems. We find clusters for ignorant and lazy-informed attack models with relatively equal subgroup bias for both systems, i.e. markers are close to identity. However, subgroup bias significantly differs in the semi-informed attack scenario with no system fully eliminating disparities. Given that the semi-informed attacker represents the strongest attack model today and the widespread distribution of subgroup bias, we advocate for fine-grained evaluation metrics acknowledging differences in subgroup performance. 
 
\subsection{Speech Recognition}
\label{sssec:subsubhead}

\begin{figure}[!t]
    \centering
    \resizebox{0.9\linewidth}{0.9\linewidth}{\input{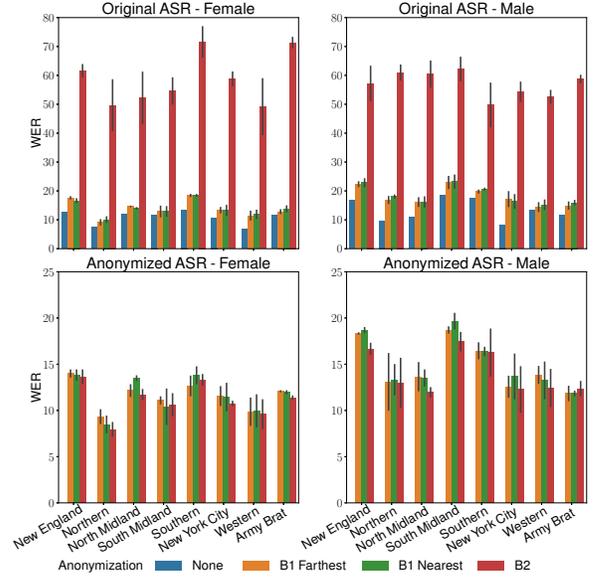}}
    \caption{ASR results for subgroups divided by dialect and sex. We show results for anonymized speech for the original ASR and an ASR trained on anonymized data.}
    \label{fig:wer}
\end{figure}

We conduct a bias analysis for speech recognition by examining sex and dialect-based differences for the three anonymization systems and attack models. Figure~\ref{fig:wer} shows the mean WER from the two benchmark datasets and their 95\% confidence intervals. Our results cover the original ASR evaluated with non-anonymized speech, the original ASR with anonymized speech and the ASR trained on anonymized speech, evaluated only on anonymized speech. 

While B2 showed similar privacy-preserving behaviour to B1.b, utility drops drastically for B2 when evaluated on the original ASR. Despite that, we find that WER differs significantly for speaker subgroups with little variance in the two benchmark variations. Male speakers show higher WER for B1.b anonymization, a possible interplay of ASR performance and slightly higher protection (see Section~\ref{sec:biasasv}). Nevertheless, disparities are consistent across ASR systems indicating that the anonymization systems maintain downstream task-relevant characteristics most likely captured by the bottleneck features. Thus, enhancing the ASR system will consequently decrease disparities in anonymized speech. However, it also highlights that the ASR system could potentially disclose subgroup membership, revealing sensitive attributes like the speaker's ethnicity to attackers.

\section{Discussion and Future Work}
\label{sec:discussion}

In this study, we conducted a comprehensive study of voice anonymization system bias to discover disparities and inform the development of anonymization techniques. We constructed a novel benchmark dataset that can be generally applied to systems that jointly implement ASR and ASV. However, our benchmark dataset was limited in size and subgroup representation resulting in reduced robustness for small subgroups. Therefore, we urge the need for larger and more inclusive benchmarks to allow robust bias evaluation. 

While our findings indicate that performance for subgroups differed significantly, they offer tangible insights for improvements such as adjustments to the pseudo-speaker pool, highlighting the importance of bias evaluation in the field. We emphasize that standard metrics like gender-dependent EER are not sufficient. Future research could apply privacy metrics like ZEBRA to investigate bias in voice anonymization considering average and worst-case protection~\cite{nautsch20_interspeech}.

We show that disparities of subgroups are consistent across ASR systems leaving room for attackers to deploy property inference attacks on the anonymized speech signals~\cite{feng2023review}. Future work could investigate new attack models that exploit biased outcomes of these systems. 
Finally, voice anonymization bias evaluation offers a unique opportunity to advance bias research in both ASR and ASV while exploring the interplay of multiple data-driven systems. Here, our benchmark dataset can prove beneficial for the broader community.
Crucially, bias assessment is an important checkpoint to make privacy-preserving voice systems more trustworthy and deployable in the real world, ensuring equal protection.

\balance
\printbibliography[heading=bibnumbered, title=REFERENCES]

\end{document}